# Non-volatile Reconfigurable Digital Optical Diffractive Neural Network Based on Phase Change Material

*Chu Wu,1,4 Jingyu Zhao, 1,4 Qiaomu Hu,1,2 Rui Zeng,1,2 and Minming Zhang1,2,3,\**

1 School of Optical and Electronic Information, Huazhong University of Science and Technology, Wuhan, Hubei 430074, China
2 National Engineering Research Center for Next Generation Internet Access System, Wuhan, Hubei 430074, China
3 Wuhan National Laboratory for Optoelectronics, Wuhan, Hubei 430074, China
4 These authors contributed equally to this work
*\* mmz@hust.edu.cn*

**Abstract**

Optical diffractive neural networks have triggered extensive research with their low power consumption and high speed in image processing. In this work, we propose a reconfigurable digital all-optical diffractive neural network (R-ODNN) structure. The optical neurons are built with $Sb_2Se_3$ phase-change material, making our network reconfigurable, digital, and non-volatile. Using three digital diffractive layers with 14,400 neurons on each and 10 photodetectors connected to a resistor network, our model achieves 94.46% accuracy for handwritten digit recognition. We also performed full-vector simulations and discussed the impact of errors to demonstrate the feasibility and robustness of the R-ODNN.

## 1. Introduction

Deep learning is a machine learning method that enables computers to finish complex tasks by simulating an artificial neural network (ANN) [1]. It has recently drastically impacted data processing for its superior performance over traditional methods. The method has widespread applications like image recognition[2], automatic driving, signal processing, and natural language processing[3]. However, electronic neural network implementations face difficulties like high energy consumption and processing time[4], [5].

Diffractive deep neural network( $D^2NN$）, has attracted extensive research for its low-cost scalability and structural simplicity. Implementations with 3D-printed diffraction surfaces and terahertz sources have achieved satisfying results on tasks like image recognition[6]–[8]. Metasurface-based $D^2$[9], [10] and on-chip integration. After the fabrication of the model, only illuminating light and photodetectors would consume power. Therefore, these designs are superior in power efficiency over traditional electronic neural networks[7], [10]. Given that $D^2$NNs use subwavelength neurons[11], [12], so migrating the design to shorter wavelengths, like the more accessible near-infrared or visible light, is a way to make the architecture more compact. However, the feature size of neurons would be so small in this situation that manufacturing becomes a significant problem. For example, a 3D printed-$D^2$NN that modulates the phase by surface height requires controlling the height of each neuron at the nanometer scale, which would be nearly impossible for existing processes. Also, the diffractive neural networks are vulnerable to misalignment and manufacturing errors[13]. To make $D^2$NN reconfigurable, researchers build active diffractive networks with devices including spatial light modulators (SLM), digital micromirror devices (DMD)[14], or reprogrammable metasurfaces[11]. These designs have achieved outstanding reconfigurability and accuracy. However, the active layers modulating light are volatile and consume much more power than the passive designs[14], [15]. The SLM-based schemes also face difficulty in integrating because of their large size.

In this work, we propose a phase-change material-powered reconfigurable digital all-optical diffractive neural network (R-ODNN). Phase-change material (PCM) based schemes provide a robust platform with convenient regulation and non-volatile modulation[16], [17]. Our network is reprogrammable to cope with different image recognition tasks without remanufacturing its structure. The digital states of our optical neurons correspond with the crystalline and amorphous states of the phase-change material $Sb_2Se_3$[18]. To fit the characteristics of the material, the R-ODNN is digitalized[19]. Since digital devices cannot train with gradients, we use a straight-through estimating (STE) method to solve the derivative problem. Featuring three diffracting layers with 14,400 neurons on each layer and ten photodetectors as output, our model achieves 93.8% recognition accuracy for MNIST handwritten digit dataset. The network's performance can boost to 94.46% with a correcting resistor network, which also dramatically improves the resilience of our network to misalignment errors.

## 2. Design

Figure 1(a) shows the structure of the R-ODNN. It consists of three digital diffractive layers, one output layer with ten photodetectors and a correcting layer. Each layer is 120μm×120μm in size and keeps a 50μm distance from neighbor layers. A diffractive layer consists of 120×120 digital optical neurons. The wavelength of the incident light is 1.55μm. At this wavelength, the refractive indices of the $Sb_2Se_3$ phase change material in the crystalline and amorphous states are 3.28 and 4.05, respectively[18], [20]. The correcting layer is a fully-connected layer with 10 inputs and 10 outputs. Figure 1(b) shows an equivalent neural network structure consisting of 3 hidden layers, a photodetector layer, and an fully-connected output layer. These layers simulate the phase change material diffractive layers, photodetectors, and the correcting layer, respectively.

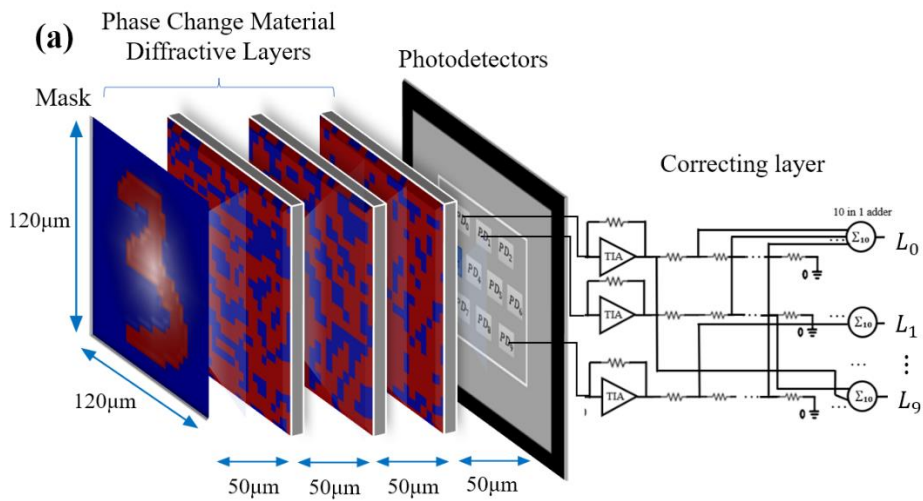

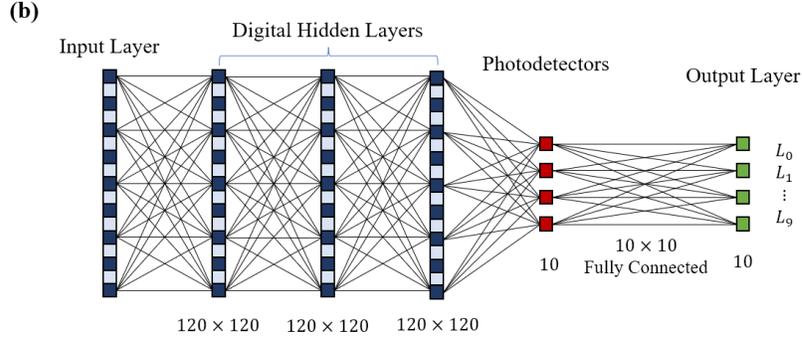

Fig. 1. Schematics of the R-ODNN. (a) The physical structure of the R-ODNN. (b) The design of the equivalent neural network.

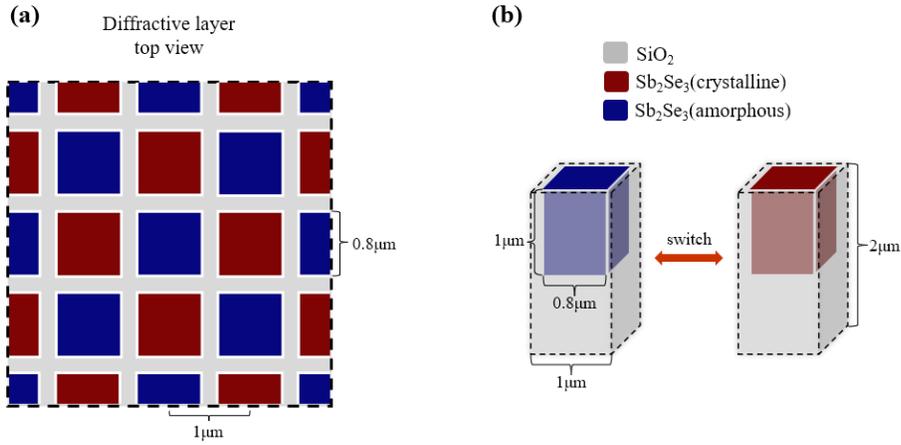

Fig. 2. Structure of the optical neurons on the diffractive layer. (a) Partial top view of the diffractive layer. (b) Single neuron structure and parameters.

Figure 2(b) shows the basic structure of the optical neuron. Optical phase-change material fills rectangular etching on the silicon dioxide substrate. The etching hole is 1μm deep and has a side length of 0.8μm, placed in grids of 1μm×1μm on the diffractive layer. We introduce parameters Θ and $K$ describe the neurons' behavior in the equivalent neural network. Θ is the phase shift difference between crystalline and amorphous states, which is $\pi$ in our case. $K$ is the ratio of the transmittance of two neuron states and is set to 1.

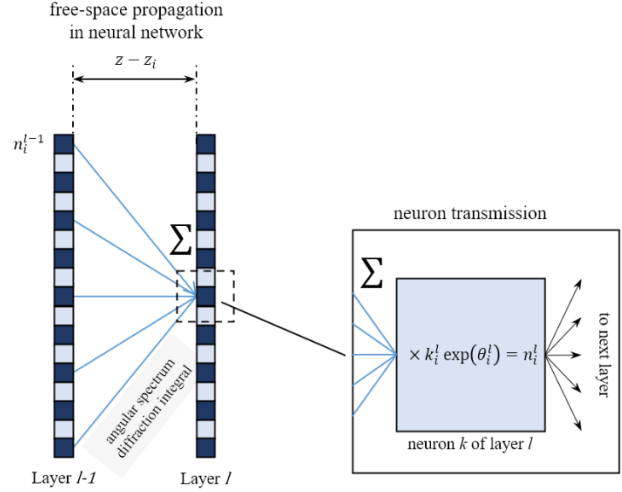

Fig. 3. Forward propagation and neuron transmission function in the equivalent neural network

The forward propagation process of the layers in the equivalent neural network is shown in Fig. 3. When the light passes through a neuron, it applies different amplitude and phase shifts to the injected light, according to Eq. 1 and 2.

$$n_i^l = m_i^l T(k_i^l, \theta_i^l) \tag{1}$$

$$T(k_i^l, \theta_i^l) = k_i^l \exp(j\theta_i^l) \tag{2}$$

$T(k_i^l, \theta_i^l)$ is the neuron transmission function of our model, where $\theta_i^l$ stands for the phase shift of the $i$-th neuron on the $l$-th layer, $k_i^l$ is the amplitude modulation of that neuron, and $m_i^l$ is the optical input of the neuron. In our network, $\theta_i^l$ is the trainable neuron weight and $k_i^l$ keeps 1. When the light propagates between layers, the diffraction pattern is calculated with angular spectrum diffraction theory[21], [22], and the input of the next layer can be told:

$$H(f_x, f_y) = e^{ikz\sqrt{1-(\lambda f_x)^2-(\lambda f_y)^2}} \tag{3}$$

$$m_i^{l+1}(x,y) = n_i^l(x',y') * F^{-1}[H(f_x, f_y)] \tag{4}$$

Here, $H(f_x, f_y)$ is the frequency-domain transmission function of diffraction, and $F^{-1}$ is the inverse Fourier transform. This procedure calculates the input $m_i^{l+1}$ of the next layer, which is the input of the neuron transmission function given in Eq. 2. When the light reaches the photodetectors on the final layer, the detectors read the intensity of the diffraction pattern and obtain classification results.

The cascaded correcting layer is a fully-connected layer that further boosts the accuracy and compensates for system errors. We may use a resistor network or digital signal processor in the physical model to finish the task. The output of the optical option, a 10-element vector $X$, is remapped through this layer into a vector of the same size. $L$ is the final output of the entire network. Equation 5 describes the mapping process.

$$L = WX \tag{5}$$

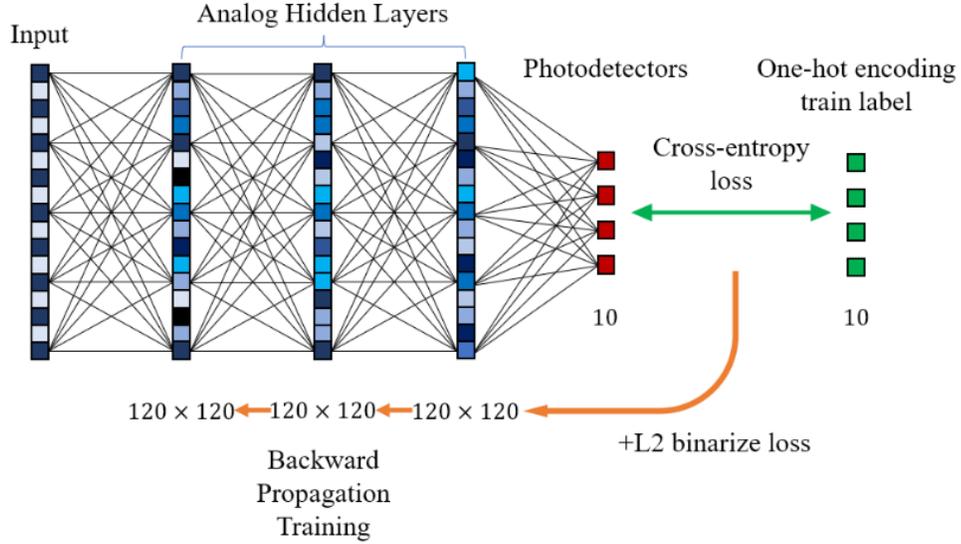

Fig. 4. The training process of the optical section of the equivalent neural network

The training process of the optical network uses backward propagation methods, as Fig. 4 shows. The output of the optical section is a 10-element vector *X*, which corresponds to the readings of photodetectors. The results use softmax normalization. Then we may calculate the loss of the network with the cross-entropy function as follows.

$$Y_i = \frac{X_i}{\sum_{j=1}^{10} \exp(X_j)} \tag{6}$$

$$L_{CELoss} = \sum_{i=0}^{9} I_i \log_2 Y_i \tag{7}$$

$I_i$ is the picture's label in one-hot encoding, and $Y_i$ is the output of the softmax function, *i* represents the *i*-th category of the label or result. The index of the photodetector with the highest output voltage is the result for classification.

Since the phase-change material only has two stable states, the transmission parameter of the neurons should also be binary. In our network, we do not perform the binarization process during the forward propagation but use a penalty function to make the parameters gather around the desired binary value. The penalty function is the sum of $l_2$-norms of the difference between a real weight and two target binary values, 0 and Θ.

$$P(\theta_i^l) = \Gamma \times (\|\theta_i^l\| + \|\theta_i^l - \Theta\|) \tag{8}$$

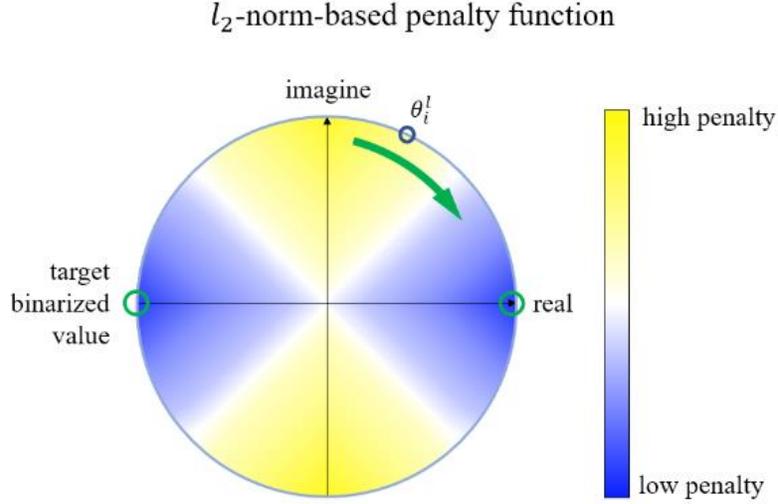

Fig. 5 The $l_2$-norm-based penalty function. Weight $\theta_i^l$ is a trainable parameter of the neuron. Penalty function forces $\theta_i^l$ moves towards the target binary value 0 or Θ at the rate defined by Γ. Here, the binary target values are 0, π respectively.

In our network, the function in Eq. 8 gives a larger penalty with input values far from 0 and π. As Fig. 5 shows, the neuron transmission function returns a complex number with phase $\theta_i^l$ and modulus 1. The penalty forces the phase $\theta_i^l$ move towards the two target binarized value, 0 and π, as the green arrow on Fig. 5 shows.

$$L_{loss} = L_{CELoss} + \sum_i \sum_l P(\theta_i^l) \qquad (9)$$

Eq. 10 is the final loss function. During the training process, the penalty function offers an additional force other than the Adam optimizer, pushing the neuron weights towards or away from the binarizing target value, 0 and π. Penalty coefficient Γ controls the direction and magnitude of this force. To achieve minimum binarization loss, we first use a negative Γ value to gather the neuron weights around binarizing thresholds, $\frac{\Theta}{2}$ and $-\left(\pi - \frac{\Theta}{2}\right)$. This method offers us a better initial value for the subsequent training process. Later, we re-set the Γ value to a small positive value and increase it along the training. In this way, the weights are gradually forced to gather around target values, 0 and Θ.

After training, we apply a binarizing function to all neuron weights. $\theta_{i,b}^l$ and $\theta_i^l$ are to represent the binary and real-valued transmission parameters, respectively. Their relationship is described in Eq. 10.

$$\theta_{i,b}^l = \begin{cases} 0, & -\left(\pi - \frac{\Theta}{2}\right) < \theta_i^l \leq \frac{\Theta}{2} \\ \Theta, & \text{else} \end{cases} \qquad (10)$$

After the binarization process, the weights $\theta_{i,b}^l$ are corresponding to the final result of the optical neuron state.

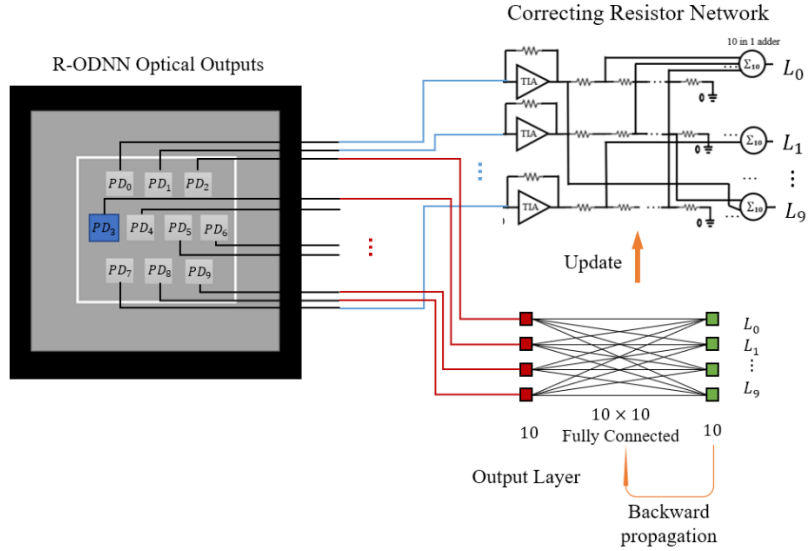

Fig. 6. The physical structure and the schematics of the correcting resistor network. The fully-connected output layer determines the values of the resistors.

After setting the optical section, we cascade the corrective section to the network and train it separately. As Fig. 6 shows, the output current of each photodetector is the input of the corrective section, and the production of adders tells the final classification result of the neural network.

We also optimize this section of the network using the softmax normalization and cross-entropy loss function to calculate the loss and the Adam method to optimize the weights of the fully-connected layer.

In corresponding to our physical model, we first train the optical section and apply the result to the optical neurons. A cascaded fully-connected layer would bring significant performance uplift for the network, especially under different error types[13]. In our case, the corrective section takes the role. Fig. 7 shows the accuracy of the R-ODNN during the training of the optical section and corrective section. The first 300 epochs show the accuracy variation as the penalty coefficient $\Gamma$ varies. The subsequent 200 epochs are the final phase of training after $\Gamma$ is set to maximum. The final epochs after Epoch 500 are the training process of the correcting section. The initial values of the correcting section are randomly set so that the accuracy drops during the first few epochs.

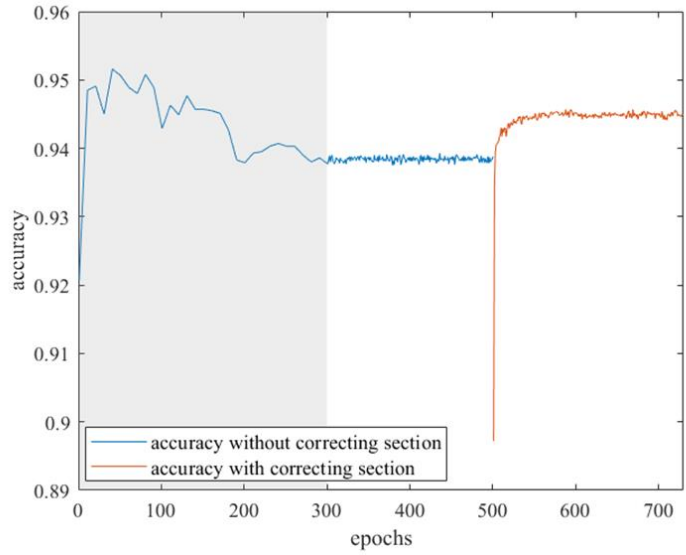

Fig. 7. Measured network accuracy with and without correcting the resistor network. The first 300 epochs use a dynamic penalty coefficient to achieve binarized weights.

## 3. Results and discussion

To further verify the feasibility and accuracy of our neural network, we perform a 3D finite-difference time-domain (FDTD) full-vector simulation via commercial software (Lumerical FDTD Solutions) for a miniature version of the model. In the neural network, the phase shift of each optical neuron is directly substituted into the diffraction integral for a layer-by-layer calculation to obtain the output. For vector simulations, the value of optical neurons is transformed into the refractive index distribution of the $i$-th diffractive layer, and then the desired diffractive pattern is obtained by FDTD simulation. We calculated the diffractive pattern of each layer using the FDTD method. Figure 8 gives an example.

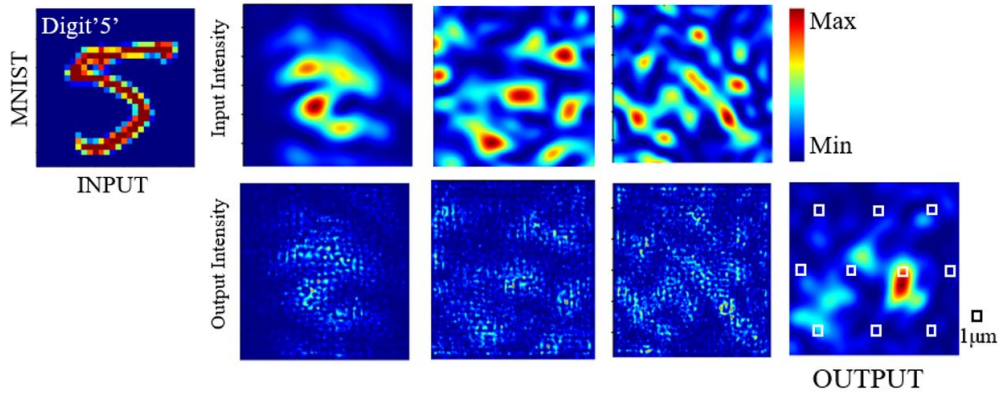

Fig. 8. The intensity of light and index distribution of each layer. The output diffraction pattern and the position of photodetectors are shown on the right.

With a similar method, we calculated several output diffractive patterns of the system. As shown in Figure 9, the diffraction patterns calculated by different methods match very well in most situations. However, minor hot spots in our final diffraction patterns could interfere with our classification results. These problems can be solved by training the correcting network online.

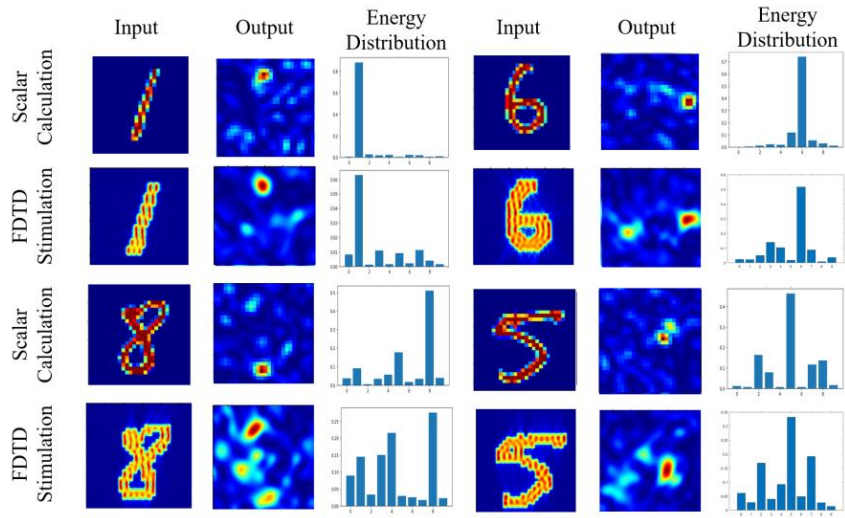

Fig. 9. The results of neural network calculation and FDTD vector simulation for handwritten digit classification. The energy distribution simulated by the two methods matches well in most situations.

According to Fig. 9, the result of our scalar diffraction theory-based neural network simulates the actual situation well in general. Therefore, we can approximate the accuracy of our physical model with the neural network. The accuracy is shown in Fig. 10, showing that the correcting network included version reached 94.5% in the end, and the all-optical model achieved an accuracy of 93.6%.

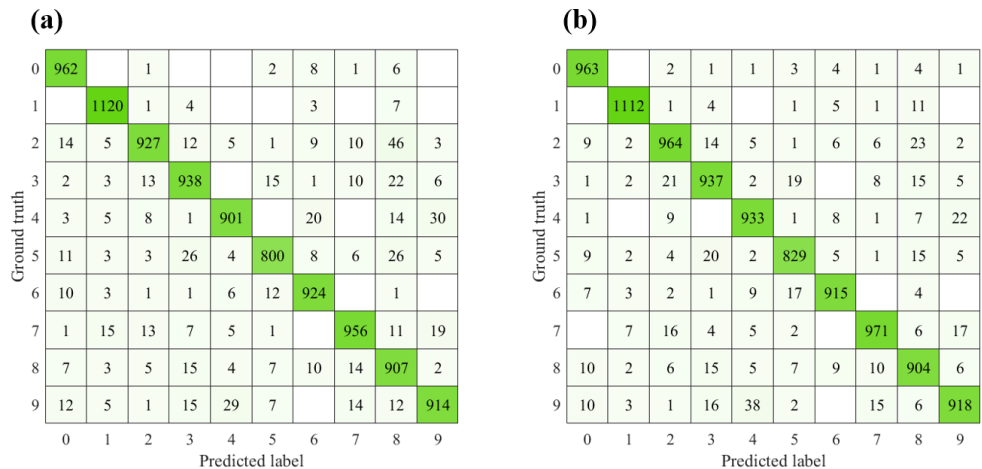

Fig. 10. The confusion matrix of the neural network after 200 epochs of training. (a) accuracy without the correcting layer; (b) accuracy with the correcting layer.

We trained the model with layer distances set at 30, 40, 45, 50, 55, 60, 65, and 70μm. As Fig. 11 shows, the classification accuracy does not change much as the layer distance varies. Considering the assembly process and the stability of accuracy, we choose a 50-55μm layer distance for our model.

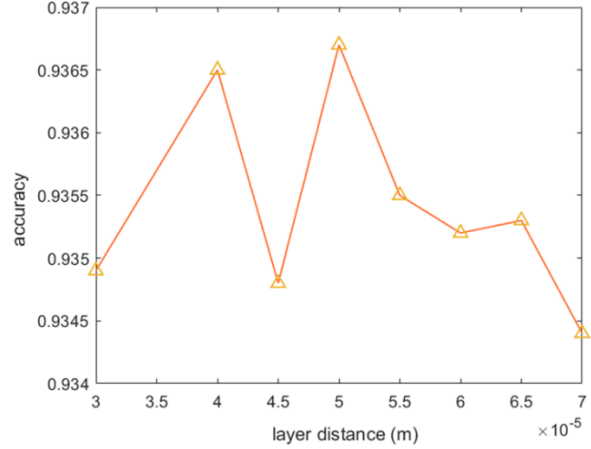

Fig. 11. The accuracy of the system with different numbers of layers and layer distance.

We apply different magnitudes of errors to the neuron, including the PCM's radius, thickness, and index to investigate the performance of our optical neuron. Figure 12 shows the variation of phase difference and the ratio of transmittance between the two PCM states when the characteristics of the neuron change.

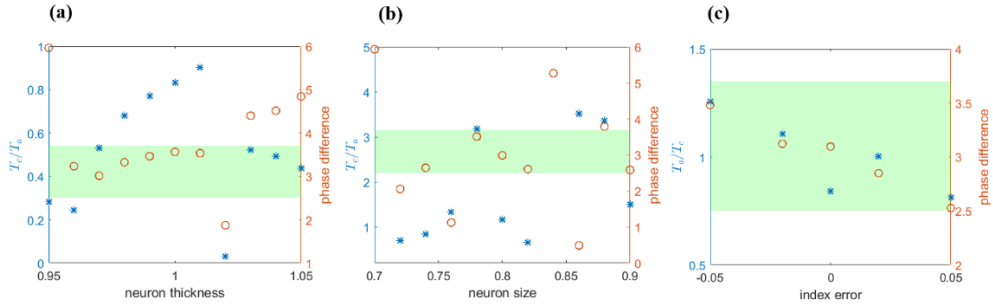

Fig. 12. The phase shift difference and the ratio of transmittance between two states of the neuron upon different types of error. The phase shift difference is the phase difference of the output light field between two states of the neuron. $T_c$ and $T_a$ are the transmittances of the neuron in the crystalline and amorphous states, respectively. The ideal value of the phase shift difference is $\pi$, and the ratio of transmittances is 1. The green area indicates the ±20% phase shift difference error around $\pi$. (a) Measured neuron performance with different neuron thicknesses, from 0.95μm to 1.05μm. (b) Measured neuron performance with different side lengths of the neuron, from 0.7μm to 0.9μm. (c) Measured neuron performance with index shift errors. The refractive index of the PCMs may deviate from the theoretical value in practice. The index shift error is measured by the deviation from the ideal index of the neuron, from -0.05 to +0.05.

Moreover, we discuss two types of manufacturing errors, assembly error, and neuron error, to further evaluate the robustness of our R-ODNN. The assembly error is the error caused by the disposition of diffractive layers. In the R-ODNN, the position of each diffraction plate in space is determined by the model's parameters, and the accuracy decreases when the diffraction layers are misspaced along the optical axis.

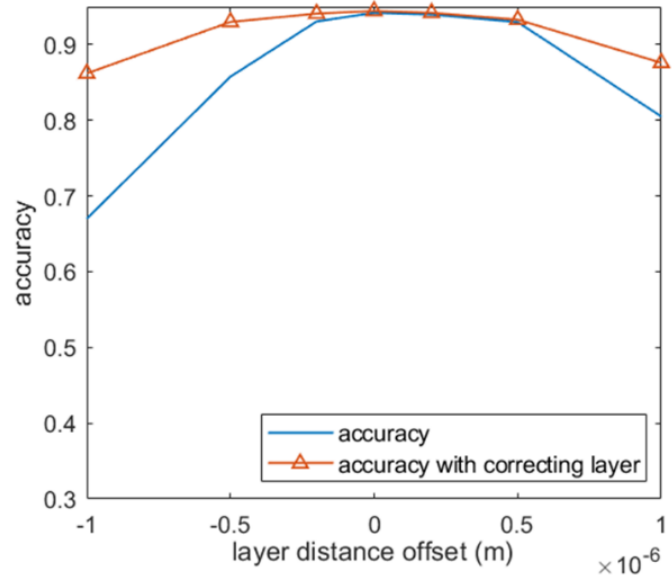

Fig. 13. The accuracy degradation of the R-ODNN under layer distance error. The orange line shows the restoration of performance brought by the correcting layer.

Figure 13 shows that R-ODNN is sensitive to the optical axis misalignment, and a severe accuracy degradation occurs at a deviation of about 1μm. However, the correcting layer we introduced earlier can be trained online, so we can retrain the correcting layer to compensate for errors after the optical structure of the R-ODNN has been fabricated. At this point, the R-ODNN can maintain an accuracy above 85% over a large tolerance.

The second manufacturing error we consider is the error in the optical characteristics of the neurons, including the transmittance and the phase shift difference. In our model, each neuron is described by two parameters $\Theta$ and $K$. Such characteristics are determined by the neuron's side length, thickness, and refractive index. Figure 13 shows the variation of phase difference and the ratio of transmittance between the two PCM states when the characteristics of the neuron change. Figure 14 shows the decrease in accuracy along with phase difference and transmittance error.

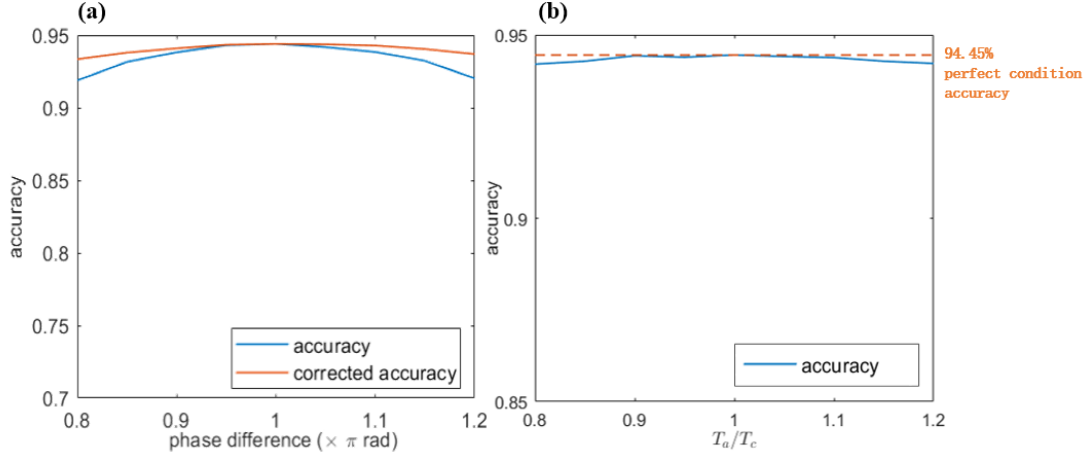

Fig. 14. Degradation of network accuracy due to the error of the phase shift difference and the ratio of transmittances. The orange curve shows the accuracy after retraining the correcting layer. (a) Accuracy of the R-ODNN when the phase difference of neurons shifts from $0.8\pi$ to $1.2\pi$, respectively. (b) Accuracy of the R-ODNN when the transmittance of two neuron states is different. The orange dashed line is the 94.45% baseline classification accuracy

The simulated accuracy of the R-ODNN drops little when the error is set within ±20%. Retraining the correcting layer online can make the degradation caused by errors even lower. According to Fig. 14(b), we can tell that the network accuracy hardly drops when the ratio of transmittances varies from 0.8 to 1.2. After retraining the correcting network, the accuracy can be further stabilized to above 85%. Altogether, our network has a high tolerance for optical neuron errors.

We made an additional simulation on the Fashion-MNIST dataset to verify the reconfigurable ability of our R-ODNN. It achieved an accuracy of 84.5% for the corrective-section-included setup, and 83.7% for an optical-only setup. The R-ODNN can do this by just reconfiguring the state of optical neurons and the values of correcting layer.

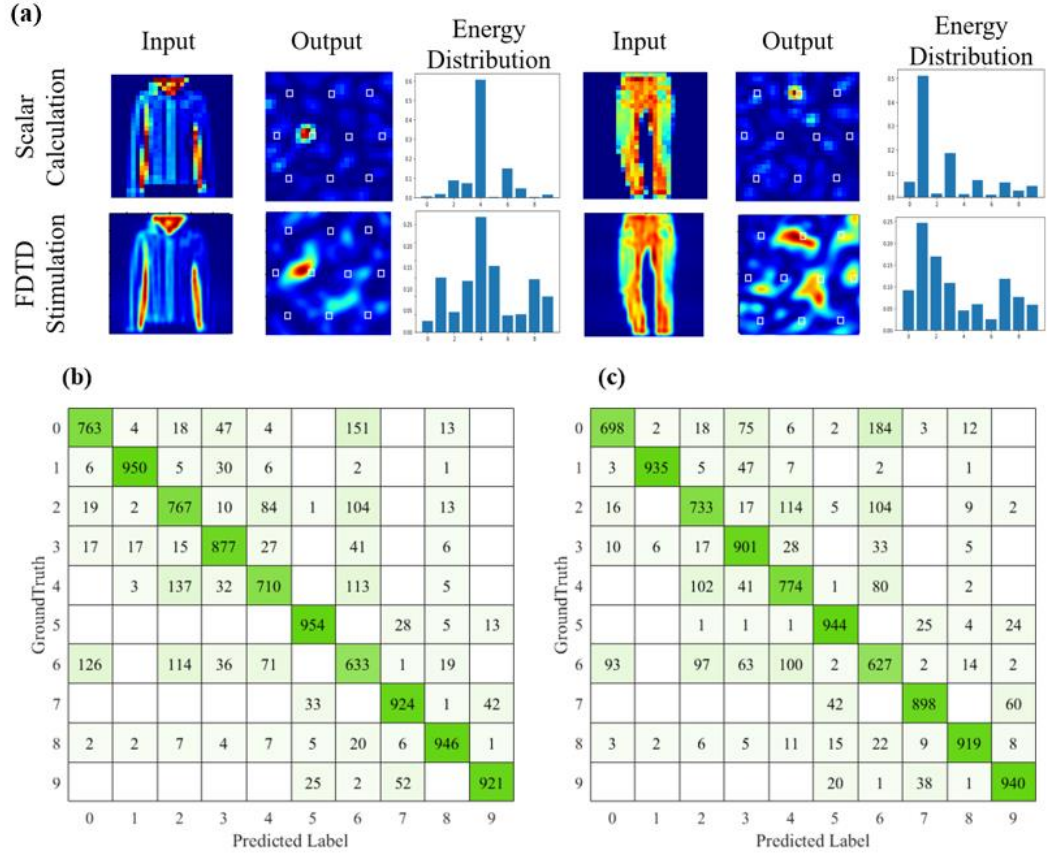

Fig. 15. The results of neural network calculation and FDTD vector simulation on the Fashion-MNIST dataset. (a) diffraction pattern of neural network calculation and FDTD simulation. (b) Confusion matrix of the complete R-ODNN setup trained with the Fashion-MNIST dataset. (c) Confusion matrix of the optical-section-only R-ODNN setup trained with the Fashion-MNIST dataset.

In Fig.15 (b) and (c), we can see the error of the label 0, 4, and 6 are significantly higher. This is probably because the Fashion-MNIST image classification dataset is more complex than the MNIST dataset. Since the size of the current optical network is relatively small, the accuracy could drop when facing complex images. Adding more optical layers or increasing the size of each layer would further improve the classification accuracy.

## 4. Conclusion

This paper proposed a reconfigurable optical diffractive neural network (R-ODNN) structure. The model includes three digital diffractive layers that each feature 1μm thickness, where digital neurons are placed on a 1μm-by-1μm grid with a side length of 0.8μm. With the resistive correcting network, the R-ODNN achieves 94.45% classification accuracy for the MNIST handwritten digit dataset and keeps above 90% under the following errors: optical axis alignment error below 1μm, layer spacing error below 2%, neuron edge length error below 5%, thickness error below 15% or refractive index error of phase change material below 10%, proving that our model is tolerant towards manufacturing errors.

The novel binarized neural network discussed in the paper reveals that simple and linear structures can perform well in optical systems but also elucidates the principal basis of an easily reconfigurable, off-line trainable, and on-chip integrated all-optical neural network device based on phase change materials. Introducing non-linear optical layers could bring significant accuracy improvement to our network. Therefore, research on non-linearity would be the topic of our future work. We also look forward to obtaining explicit performance in the subsequent physical fabrication experiments.

**Funding.** National Natural Science Foundation of China (62175076; 61775069; 51911530159).

**Disclosures.** The authors declare no conflicts of interest.

**Data availability.** Data underlying the results presented in this paper are not publicly available at this time but may be obtained from the authors upon reasonable request.